\documentstyle[12pt]{article}
\begin{document}
\begin{titlepage}

\hfill{PURD-TH-92-10}

\hfill{UM-P-92/70}

\hfill{OZ-92/22}

\vskip 1cm

\centerline{\Large \bf Discrete quark-lepton symmetry need not pose}
\vskip 2 mm
\centerline{\Large \bf a cosmological domain wall problem}

\vskip 1.5cm

\centerline{H. Lew$^{(a)}$\footnote{email: lew@purdd.hepnet,
lew\%purdd.hepnet@LBL.Gov} and R. R. Volkas$^{(b)}$\footnote{email:
U6409503@ucsvc.ucs.unimelb.edu.au}}

\vskip 1.5cm

\noindent
{\it (a) Physics Department, Purdue University, West Lafayette, IN
47907-1396, U.S.A.}

\vskip 2 mm
\noindent
{\it (b) Research Centre for High Energy Physics, School of Physics,
University of Melbourne, Parkville 3052, Australia}

\vskip 1.5 cm

\centerline{ABSTRACT}

\noindent
Quarks and leptons may be related to each other through a spontaneously
broken discrete symmetry. Models with acceptable and interesting
collider phenomenology have been constructed which incorporate this
idea. However, the standard Hot Big Bang model of cosmology
is generally considered to eschew
spontaneously broken discrete symmetries because they often lead to the
formation of unacceptably massive domain walls. We point out that there
are a number of plausible quark-lepton symmetric models which do {\it not}
produce cosmologically troublesome domain walls. We also raise what we
think are some interesting questions concerning anomalous discrete
symmetries.

\end{titlepage}

\leftline{\bf 1. Introduction}
\vskip 5mm

In the early 1960s, a disconcerting imbalance in the spectrum of
quarks and leptons was uncovered. With the discovery that the
muon-neutrino was a distinct flavour it seemed that there were four
fundamental leptons, but only three quarks. Largely on the basis of
aesthetics, several people speculated that this asymmetry would
eventually be rectified by the discovery of a fourth
quark\cite{charm}. Their sense
of aesthetics was vindicated in the mid-1970s with the experimental
detection of charm. The idea that quarks and leptons are paired up in
each fermion generation is now a familiar and pleasing fact of nature.

Although the charm quark was introduced on the basis of a desired
``symmetry'' between quarks and leptons, there really is no symmetry
in the rigorous sense of the word between these fermions in the Standard
Model (SM). Quarks are
coloured; leptons are not. Leptons have integral charge; quarks have
not. Quark and lepton masses are quite different.
Furthermore, there is no definitive evidence for the existence of the
right-handed neutrino, which is the putative partner of the right-handed
up quark. Does all of this
mean that the successful aesthetic of the 1960s is in truth only
partially adhered to?

The answer is actually ``that we do not know'' rather than a
loud ``no''. Recently it has become clear that quarks and leptons
{\it might} be more closely related to each other than is currently
evident. Furthermore, evidence for such a relationship could be
uncovered at energy scales as low as a few hundred GeV. This represents
an attractive confluence between theoretical speculation and hard-core
phenomenology.

This speculative relationship between quarks and leptons
involves the ideas of ``leptonic colour'' and ``discrete quark-lepton
(q-$\ell$) symmetry''\cite{flql}.
It is a gauge-theoretic fact that the leptons we observe
might be just the lightest components of triplets under a spontaneously
broken SU(3) gauge symmetry for leptons. This leptonic colour group, if
it exists, would nicely reflect the attributes of its quark cousin.
Quarks and leptons would appear much more like each other than they do
in the Standard Model.

But having gone to the trouble of introducing a spontaneously broken
leptonic colour group, it is very tempting to push the quark-lepton
association still further by postulating that a rigorous, but
spontaneously broken, discrete symmetry
exist between the two sectors. This would be the logical culmination of
the primordial aesthetic which lead to the
experimentally vindicated hypothesis of charm.
Nature may or may not
make use of leptonic colour or discrete quark-lepton symmetry.
But, it is surely very interesting to find out.\footnote{It is also
interesting to wonder about why the quark-lepton symmetry idea took so
long to be proposed. A possible reason is that grand unified theories
(GUTs) were proposed very soon after the wide-spread acceptance of the
SM in the early 1970s. This alternative way of connecting the quarks and
the leptons became, and still is, very influential, and so people may
have felt that nothing qualitatively different from this was possible.}

Models with q-$\ell$ symmetry yield a rich non-standard
phenomenology\cite{flvph}: exotic charge
$\pm$1/2 fermions (liptons) confined by a new unbroken
asymptotically free SU(2) gauge interaction, light exotic SU(2) glueball
states\cite{flvglue},
new heavy gauge bosons and a number of new Higgs bosons. Since much of
this new phenomenology is allowed to exist in the 100 GeV to 1 TeV
energy regime, q-$\ell$ symmetric models should be of great interest to
the phenomenologists and experimentalists of today.

Despite the appeal of discrete q-$\ell$ symmetry from a purely particle
physics perspective, aficionados of the now standard Hot Big Bang Model
(HBBM) of cosmology are likely to be less than enthusiastic about it,
for reasons we will now explain.

In its simplest form,
discrete q-$\ell$ symmetry is isomorphic to the group $Z_2$.
When a $Z_2$ discrete symmetry spontaneously breaks, the vacuum
manifold consists of two
disconnected pieces which are related by a $Z_2$ transformation.
If we ignore all of the other isometries of this manifold, then we can
consider it to consist of only two
(isolated) states which can be transformed
into each other by the discrete symmetry. Since the
actual vacuum state in a causally connected region of spacetime has to
evolve into a unique state after a suitable relaxation time,
one of these two candidate vacua is selected as the actual vacuum.
A cosmological problem arises here, however, because
spacetime immediately after the Big Bang consists of causally
{\it disconnected} regions. This means that, at the time of the
cosmological phase transition associated with the spontaneous breaking
of the discrete symmetry, randomly different choices for the
actual vacuum will in general be made in the various causally
disconnected pieces of spacetime. But, as the universe
continues to expand after this phase transition,
regions that previously had no influence over each other come
into causal contact. Two such regions that happen to have
different vacuum states therefore have to
form a domain wall structure at their boundary, if there is insufficient
energy to flip the vacuum state in one the domains.
(Although our discussion
was restricted to the $Z_2$ case for simplicity, the existence of domain
walls follows for all discrete symmetries.)

This reasoning is born out by examining the classical solutions
of field theories which display spontaneously broken discrete symmetries,
because they include solutions describing topologically stable
domain walls. Using
these classical solutions, one can calculate the energy per unit area of
a domain wall and hence conclude that such structures
should dominate the energy content of the observable universe
(unless the scale of
discrete symmetry breaking is much less than the electroweak scale).
Since this is contrary to observation, theories predicting stable domain
walls are inconsistent with the HBBM of cosmology\cite{zel}.
The purpose of this paper is to show that certain classes of q-$\ell$
symmetric theories evade this potential problem.

The conclusion that domain walls are a cosmological calamity relies on a
number of assumptions: (i) that the HBBM is the correct model to use;
(ii) that the domain walls are stable;
(iii) that there is a cosmological phase transition associated with the
spontaneous breaking of the discrete symmetry (in other words, that there
exists a critical temperature $T_c$ above which the discrete symmetry is
restored); (iv) that an inflationary period did not occur after the
discrete symmetry phase transition; and (v) that the two states in the
vacuum manifold are really degenerate. There may also be other
important assumptions that we have failed to notice.

In the remainder of this paper we will discuss each of these five
assumptions with special emphasis on their role in determining the
cosmological consequences of
spontaneously broken q-$\ell$ symmetry. In Section 2 we discuss the
status of the HBBM and its connection with particle physics [see
assumption (i) above]. This puts
into perspective the analysis that is to follow. Section 3 is devoted to
a brief review of the minimal q-$\ell$ symmetric model.
We then go on to show in Sections 4, 5 and 6 that assumptions
(ii), (iii) and (iv), respectively,
need not hold in plausible q-$\ell$ symmetric models,
thereby demonstrating the existence of cosmologically benign gauge
theories with discrete q-$\ell$ symmetry. In all of these cases, we will
emphasise that the resulting theories can yield interesting new
phenomenology in the 100 GeV to 1 TeV regime.
We will pose some interesting questions concerning assumption (v) in
Section 7, but we will not be able to answer them fully. Section 8
contains our conclusions.

\vskip 1cm
\leftline{{\bf 2. Status of the Hot Big Bang Model}\footnote{Although
the contents of this section are relevant, they do not need to be
digested in order to understand the main body of the paper. Those
readers who wish to skip this section may do so without loss of
continuity.}}
\vskip 5mm

In the last decade or so, the fields of cosmology and particle physics
have become deeply intertwined. The HBBM postulates that the universe
originated in an extremely (perhaps infinitely) hot and dense state
about 10-15 billion years ago, and that it has been expanding
quasi-statically ever since. The existence of high-temperature
equilibrium states in the very early universe ties cosmology and
particle physics together, according to this scenario. In this section,
we will try to give a balanced assessment of how important it is to
ensure that new ideas in particle physics are consistent with Big Bang
cosmology.

The HBBM enjoys some interesting observational support. Most
fundamentally, the observed red-shifts of galaxies and related objects
can be simply explained by the hypothesis of an expanding universe.
Also, a qualitative evolutionary trend with red-shift for galaxies is
evident, which dovetails neatly with the HBBM view that distant galaxies
should possess features characteristic of much earlier times in the
evolution of the universe. The microwave background, with its striking
black-body spectrum, is consistent with a very hot epoch in the early
universe. These three pieces of evidence demonstrate that an expanding
universe which was once very hot bears an interesting similarity with
our universe as revealed by observational astronomers.

Evidence for a deep consistency between the HBBM of cosmology and the SM
of particle physics comes primarily from calculations of the primordial
abundance of hydrogen, helium and lithium isotopes.
Big Bang Nucleosynthesis (BBN) predicts that the universe is
essentially three parts hydrogen to one part $^4$He, with {\it calculable}
traces of deuterium, $^3$He and $^7$Li
(the production of other isotopes being
ascribed to secondary nuclear processes within stars). The BBN
calculations depend critically on nuclear interaction rates and the
expansion rate of the universe at the time of nucleosynthesis. The
expansion rate in turn depends critically on the number of relativistic
degrees of freedom in equilibrium at this time. The SM of particle
physics says that only electrons, positrons, photons, neutrinos and
antineutrinos can be relativistic during the BBN era. The only possible
freedom in altering the number of relativistic species therefore rests
with the number of neutrino flavours. Detailed observational work on
primordial light element abundances now exists, and BBN calculations
predict the correct numbers provided that (a) the number of light
neutrino flavours is three, and (b) that the baryon density of the
universe is 5-10 times that observed in luminous astronomical bodies. As
everybody knows, the first of these necessary conditions was
dramatically confirmed by the LEP collider at CERN through the precise
measurement of the $Z$-width. Thus primordial nucleosynthesis within the
HBBM is consistent with our knowledge of particle physics.

The above discussion summarizes the well-known successes of the HBBM.
There are also some well-known theoretical problems with it\cite{linde}.
The evident
large-scale homogeneity and isotropy of the observed universe
(its ``smoothness'') is at odds
with the existence of causally-disconnected regions of spacetime in the
early universe. For instance, the HBBM asserts that the
observable universe of today
evolved from about $10^6$ causally-disconnected spacetime volumes
at the time of radiation decoupling. How then can the isotropy of the
microwave background radiation be explained?
On another tack, we know observationally that the
universe is very close to being spatially flat. However the
Einstein equations describing the expansion of the universe require very
special initial conditions in order to bring this about. In particular,
the average mass density of the universe $\rho$
must be equal to the critical
density $\rho_c$ to one part in $10^{59}$ at the Planck time.
Such a special value demands
an explanation which is not forthcoming in the HBBM.

One interesting hypothesis has been advanced to rid the HBBM of the
smoothness and flatness problems: inflation\cite{linde}.
The idea is that sometime
in the very early universe, a period of exponential expansion ensued.
This allows the present day universe to have evolved from within a
causally-connected region of spacetime, and thus the smoothness of our
universe is no longer a mystery. Also, the spatial metric after
the inflationary epoch is flat to an extremely high precision, so
ameliorating the other theoretical problem of the
unadorned HBBM.\footnote{Inflation was originally also motivated by the
monopole problem. However, since this presupposes that the U(1) gauge
group of electromagnetism is embedded in a larger group without a U(1)
factor it is not inevitable that a monopole problem arises in the first
place.}

The Inflationary Hot Big Bang Model (IHBBM) has an important prediction:
the energy density of the universe is equal to the critical density
$\rho_c$ to an extraordinarily high precision. Now, matter in luminous bodies
accounts for only about 1\% of this predicted value. If BBN is believed,
then a further 5-10\% of $\rho_c$ should exist as baryons in
non-luminous bodies (including possibly black holes). Also, the
gravitational rotation curves of spiral galaxies indicate the presence
of some dark gravitating matter of the order of about 10\% of $\rho_c$,
while the dynamics of galactic clusters furnishes some evidence of dark
matter at the level of about 20-30\% of $\rho_c$.
So, if inflation is correct, 70-90\% of the matter in the universe is
yet to be identified, either directly as luminous objects, or indirectly
through gravitational effects, or even more indirectly through its
effect on BBN calculations.

The IHBBM, with its prediction of dark matter, is popular with
theorists who are trying to understand the formation of
large scale structure (galaxies,
clusters of galaxies, clusters of clusters of galaxies
and so on). The point is that microscopic inhomogeneities
caused by quantum fluctuations in the pre-inflationary universe
get amplified during inflation into macroscopic inhomogeneities. These
could then be the seeds for large-scale clumping via the Jeans
instability. An especially popular scenario sees the clumping initiated
by inhomogeneities in cold dark matter.

The framework summarized above seems to be generally consistent with
astronomical observations. It has appeal because of its underlying
simplicity, and probably represents the ``best guess'' we have about
how our universe evolved. However, we should not be so euphoric about
its success that we lose our customary and correct
stance of scientific objectivity and continual criticism. We should also
not close the door on creative new ideas, or interesting modifications
of standard ideas. In cosmology, we face a serious problem
in regard to {\it testability}. Evidence in favour of the HBBM, though
it exists, is not so overwhelming that we should brook no alternatives
or modifications.

For instance, as we said, the unadorned HBBM has serious problems
explaining the homogeneity, isotropy and flatness of the universe. These
problems can be solved if the idea of inflation is added to the HBBM.
But how do we test inflation? Its major prediction is that the mass
density of the universe is equal to the critical value $\rho_c$
to extraordinary precision. While
this is {\it consistent} with current observational data, these data do
not allow one to make nearly as precise a test as the IHBBM really
requires. So, although inflation is an {\it interesting} idea, it is not a
{\it well-tested} idea. Furthermore, since without inflation the HBBM
has theoretical deficiences,\footnote{It is
possible there are ways other than inflation to modify the HBBM in order
to solve the smoothness and flatness problems. For instance, it has been
suggested that quantum gravity may provide a probability density for
spacetimes which is strongly peaked for smooth, symmetrical
configurations.} and since inflation is not a well-tested idea and could
still be wrong, it is sensible to spend some effort in searching
for alternatives to the HBBM framework itself.\footnote{Of course, for
alternatives to be acceptable, they would also have to be consistent
with the classic evidence used in favour of the HBBM, or they would have
to somehow reinterpret this evidence. We are not
necessarily claiming that any such alternatives are known, although an
attempt by Arp et al.\cite{arp} has been discussed in the literature.
Whether or not the arguments of Arp et al.\ hold up to scrutiny
is irrelevant to the point we are trying to make here. As far as we are
aware, there is no ``no-go theorem'' about alternative cosmological
models. Difficulty in finding a good foil for the HBBM should not be
mistaken for the non-existence of same. Maybe we are just not smart
enough, or not knowledgeable enough yet.}

Another example is furnished by BBN. In order for these calculations to
come out correctly, we need to postulate
several times the number density of
baryons that we can readily account for in luminous bodies. Until this
baryonic dark matter is found, there is a serious loose end in BBN. An
intriguing {\it possibility} is that the baryonic
dark matter may exist in
galactic haloes as ``Jupiters'' or something similar. Then maybe the
anomalous galactic rotation curves can be connected with BBN.
Astronomers are currently trying to test this idea through the
observation of ``microlensing''. However, we are not there yet. Maybe the
required baryons are located elsewhere; maybe we will not know for a long
time whether baryonic dark matter exists or not.
We have to live with this, and do the best we can.
Trying to piece together what happened in the very early universe from
the debris left around today is an extremely difficult task, and is the
main reason why the testability of cosmological models is such a
problem.

By way of contrast, particle physicists are much more fortunate.
Consequently, the standard SU(3)$_c\otimes$SU(2)$_L\otimes$U(1)$_Y$
model of particle physics has been subjected to great scrutiny and has
passed the most searching examinations with its reputation enhanced
(even at the level of radiative corrections).
There are nevertheless still some loose ends, such as direct proof that
the tau-neutrino is a distinct flavour, direct proof that top exists and
direct proof that a single, neutral physical Higgs boson exists. One
can also add the issue of neutrino masses to this list. It is remarkable
that even in this extremely well-tested
context, particle physicists continue to think
about non-standard properties for the top-quark
(for instance, top-mode symmetry breaking) and the Higgs sector
(multi--Higgs-doublet models, top-mode symmetry breaking,
technicolour and so on). It is therefore
recognized in particle physics that loose ends need to be tied up, and
that new phenomena can arise even within a very well-tested framework.

It is therefore somewhat ironic that great currency is given to
constraints on new particle physics derived by demanding that the
standard cosmological scenario not be disturbed. This is where the issue
of domain walls and spontaneously broken discrete symmetries arises.
Domain walls are members of a long list of objects\footnote{Axions,
heavy (e.g.\ 17 keV) neutrinos, monopoles, extra light degrees of
freedom, stable fractionally charged free particles ...} that need
special treatment in order to be compatible with the HBBM. How important
is it to study the details of such ``special treatment''?

We hope we
have implicitly given a fair and rational answer to this question during
this short review of cosmology and particle physics. Explicitly,
we believe that it is reasonable to study interesting extensions of
the Standard Model of particle physics, even if they are not compatible
with the Hot Big Bang Model of cosmology. The point is we can test
particle physics models in great detail in terrestrial
laboratories, provided they do not involve phenomena beyond
technologically feasible
energy scales. These sorts of theories can, with complete confidence,
be either ruled-out in the laboratory, or
relegated to being allowed only at terrestrially inacessible
(and thus relatively uninteresting) energies,
so cosmological considerations do not need to be
used to evaluate these theories.
We may unwisely dismiss some interesting and potentially
important ideas in particle physics
if we take cosmological constraints as completely rigorous. Since
cosmological models can only be tested with quite limited precision, it
is not reasonable to view cosmological constraints completely without
suspicion. On the other hand, the HBBM (or its inflationary extension) is
an impressive scenario that seems to be consistent with all available
observational data. Therefore, {\it although the compatibility of
spontaneously broken discrete q-$\ell$ symmetry with standard cosmology
is not necessary for it to be an important idea, it is nevertheless {\em
interesting} to see under what circumstances it is in fact compatible.}

Before turning to our analysis of this, we should conclude this section
by asking what we believe to be an important question: What criteria
would have to be met by a cosmological model for it to be considered
well-tested? Although we do not know the definitive answer to this, we
would suggest that a necessary condition is a detailed, predictive
theory of large scale structure formation that fits in smoothly with
known particle physics (as revealed by accelerators\footnote{And maybe
also cosmic rays}). In the foregoing, we briefly alluded to the appeal
of inflation in this regard. In this context, a well-tested theory
should contain at least (i) a verification that $\rho=\rho_c$, (ii)
correct predictions for the statistical distribution of galaxies,
clusters of galaxies, clusters of clusters of galaxies, voids, walls and
filaments,
(iii) a correct account of anisotropies in the microwave background,
(iv) experimentally verified dark matter, and (v) a well-motivated
and theoretically consistent
Higgs field to drive inflation.\footnote{Unfortunately, with the high
scales that are typically associated with inflation it may never be
possible to test ``inflaton'' Higgs sectors in the laboratory.}
The recent measurement of microwave
background anisotropies from the COBE satellite\cite{smoot}
is a pleasing development
in our study of the formation of large-scale structure. This sort of
detailed work must continue\cite{COBE}.

\vskip 1cm
\leftline{\bf 3. The minimal quark-lepton symmetric model}
\vskip 5mm

In the following we will give a brief summary of the essential
features of the minimal quark-lepton symmetric
model\cite{flql,flvph}.
This will serve as a starting point from which discussions of
other models involving q-$\ell$ symmetry are based while at the
same time establishing the notation of the paper.

The minimal gauge model illustrating the basic idea of
q-$\ell$ symmetry is obtained by enlarging the Standard Model
gauge group to $G_{q\ell}$, where
\begin{equation}
G_{q\ell} = SU(3)_{\ell} \otimes SU(3)_q \otimes SU(2)_L \otimes
U(1)_X.
\label{Gql}
\end{equation}
Here SU(3)$_q$ is the usual colour group and SU(3)$_{\ell}$ is its
leptonic partner. This enlargement
requires a tripling in the number of leptons.
Each standard lepton (the left-handed electroweak doublet $f_L$,
the right-handed charged lepton $e_R$ and the right-handed
neutrino $\nu_R$) has two exotic partners, hereafter called ``liptons''.
The expanded fermionic generation is defined by the transformation laws
\begin{eqnarray}
F_L &\sim & (3,1,2)(-1/3),\ \ E_R \sim (3,1,1)(-4/3),\ \ N_R \sim
(3,1,1)(2/3), \nonumber \\
Q_L &\sim &(1,3,2)(1/3),\ \ u_R \sim (1,3,1)(4/3),\ \ d_R \sim
(1,3,1)(-2/3).
\label{minimalf}
\end{eqnarray}
The standard lepton doublet $f_L$ is embedded in $F_L$,
$e_R$ in $E_R$ and $\nu_R$ in $N_R$.
The $Z_2$ discrete symmetry
\begin{equation}
F_L \leftrightarrow Q_L,\ E_R \leftrightarrow u_R,\ N_R
\leftrightarrow d_R,\
G^{\mu}_q \leftrightarrow G^{\mu}_{\ell},\ C^{\mu}
\leftrightarrow -C^{\mu}
\label{minimalqlsym}
\end{equation}
can now be defined [where $G^{\mu}_{q,\ell}$ are the gauge bosons
of SU(3)$_{q,\ell}$ and $C^{\mu}$
is the gauge boson of U(1)$_X$].
Standard hypercharge is identified as the linear combination $X
+ {1\over 3}T$ where
$T = {\rm diag}(-2,1,1)$ is a generator of SU(3)$_{\ell}$. Standard
leptons are identified with the $T=-2$ components of the leptonic colour
triplets, while the $T=1$ components are the charge $\pm$1/2 liptons.

In order to spontaneously break SU(3)$_{\ell}$ and the
quark-lepton discrete symmetry, and also to give
masses to the liptons, the Higgs
bosons $\chi_1$ and $\chi_2$ are introduced. These scalars are
defined through the Yukawa Lagrangian
\begin{equation}
{\cal L}^{(1)}_{Yuk} = h_1 [\overline{F_L} (F_L)^c \chi_1 +
\overline{Q_L} (Q_L)^c \chi_2]
+ h_2 [\overline{E_R} (N_R)^c \chi_1 + \overline{u_R}
(d_R)^c \chi_2] + {\rm H.c.}
\label{yuk2}
\end{equation}
where $h_{1,2}$ are the Yukawa couplings and family indices
have been suppressed. The quantum numbers of the Higgs fields, and
their behaviour under the discrete symmetry, are
\begin{equation}
\chi_1 \sim (\overline{3},1,1)(-2/3),\quad
\chi_2 \sim (1,\overline{3},1)(2/3);\qquad
\chi_1 \leftrightarrow \chi_2.
\end{equation}
The $T = 2$ component of $\chi_1$ develops a nonzero vacuum expectation
value (VEV), while the VEV of $\chi_2$ is completely zero.

Electroweak symmetry breaking is achieved through a SM Higgs
doublet, which is defined through the
analogue of the standard Yukawa Lagrangian:
\begin{equation}
{\cal L}^{(2)}_{Yuk} = \Gamma_1 (\bar F_L E_R \phi + \bar Q_L
u_R \phi^c)
+ \Gamma_2 (\bar F_L N_R \phi^c + \bar Q_L d_R \phi) + {\rm H.c.}
\label{yuk1}
\end{equation}
This Lagrangian has the same purpose as in the SM. The Higgs
field $\phi$ has quantum numbers given by
\begin{equation}
\phi \sim (1,1,2)(1).
\end{equation}
Under quark-lepton
symmetry $\phi$ has to transform into its charge conjugate field
(i.e. $\phi \leftrightarrow \phi^c$) since the U(1)$_X$ gauge field changes
sign (i.e. $C^{\mu} \rightarrow -C^{\mu}$) under the operation of the
quark-lepton discrete symmetry.

The Yukawa Lagrangian yields the tree-level mass relations,
\begin{equation}
m_u = m_e, \quad m_d =m^{\rm Dirac}_{\nu}.
\label{massrels}
\end{equation}
Here $m_{u,e,d,\nu}$ refer to the $3 \times 3$ mass matrices
($u$ refers to charge 2/3 up-like quarks, $e$ refers
to the charged leptons, etc.).
These mass relations arise as
a consequence of (i) the assumption that quark-lepton symmetry is
a symmetry of the Yukawa Lagrangian and, (ii) using
the minimal Higgs sector of only one doublet. It would be impressive if
a q-$\ell$ symmetric model could be found which contained radiative
corrections that transformed these tree-level mass relations into
correct and predictive results. No such model has as yet been
constructed, although a certain q-$\ell$ symmetric model with a
non-minimal gauge group has been
shown to contain radiative corrections which can yield correct but
unpredictive fermion masses\cite{flmass}
(indeed a further extension of this
non-minimal model will be used in the next section).
If the minimal model is extended to
contain two Higgs doublets, then the relations of Eq.~(\ref{massrels})
can be avoided at tree-level but at the expense of predictivity.
Therefore, discrete
q-$\ell$ symmetry is certainly not incompatible with the measured quark
and lepton masses.

For future reference we mention that the mass relation involving
the neutrinos can be avoided if Majorana masses are given
to the right-handed neutrinos. This can be achieved through
the Higgs multiplet $\Delta_1$ as defined in
\begin{equation}
{\cal L}^{(3)}_{Yuk} =
n\left[\overline{N_R} (N_R)^c \Delta_1
+ \overline{d_R} (d_R)^c \Delta_2 \right] + {\rm H.c.}
\label{yuk3}
\end{equation}
where
\begin{equation}
\Delta_1 \sim (6,1,1)(4/3), \quad
\Delta_2 \sim (1,6,1)(-4/3);\qquad
\Delta_1 \leftrightarrow \Delta_2.
\end{equation}
It is assumed that the $T=-4$ component of $\Delta_1$
develops a nonzero VEV while the VEV of $\Delta_2$ remains zero.

The symmetry breaking pattern can be summarised as follows:
\begin{eqnarray}
&SU(3)_{\ell} \otimes SU(3)_q  \otimes  SU(2)_L
\otimes U(1)_X &\nonumber \\
&\langle\Delta_1\rangle\ \downarrow\
\langle\chi_1 \rangle &\nonumber \\
&SU(2)' \otimes SU(3)_q \otimes SU(2)_L \otimes U(1)_Y&\nonumber \\
&\ \ \ \ \ \ \downarrow\
\langle\phi\rangle &\nonumber \\
&SU(2)' \otimes SU(3)_q \otimes U(1)_Q&
\label{pat}
\end{eqnarray}
The SU(2)$'$ is an unbroken gauge symmetry. This gauge force
is expected to be asymptotically free. In analogy with QCD,
we assume that it confines all SU(2)$'$ coloured states,
so that at large distances only colour singlet states exist
in the spectrum.

\vskip 1cm
\leftline{\bf 4. Unstable domain walls}
\vskip 5mm

We will now begin our investigation of assumptions (ii)-(v) as
identified in the Introduction.

In this section, we will discuss one way in which assumption (ii) -- that
domain walls are stable -- can be evaded in q-$\ell$ symmetric models.
The basic idea is not new: we find a way to embed the discrete symmetry
inside a continuous symmetry\cite{ev}.
We then envisage that the continuous
symmetry spontaneously breaks to the discrete symmetry at a high scale,
followed subsequently by the spontaneous breaking of the discrete symmetry
at a lower scale. During the first cosmological phase transition, a
network of cosmic strings forms. These cosmic strings then have to form
the boundaries of the domain walls produced after the second phase
transition. The dynamics of these hybrid string-wall structures is such
that the domain walls are eventually ripped apart, thus rendering
them unstable and cosmologically benign\cite{ev}.

We will take the dynamics of the string-wall structures as given\cite{ev}.
Our task is therefore to show how discrete q-$\ell$ symmetry can be
embedded in a continuous symmetry and how the two stages of spontaneous
symmetry breaking can be induced. We will also have to ensure that no
other cosmological problems are introduced in the process.

The gauge group of the minimal q-$\ell$ symmetric
model is given by $G_{q\ell}$ in Eq.~(\ref{Gql}).
However, for the purposes
of this section we have to begin with a slightly more complicated gauge
group, given by $G'_{q\ell}$ where
\begin{equation}
G'_{q\ell} = SU(3)_{\ell} \otimes SU(3)_q \otimes SU(2)_L \otimes U(1)_R
\otimes U(1)_V.
\label{Gqlpr}
\end{equation}
The group U(1)$_R$ is just the Abelian
subgroup of the (non-existent) right-handed weak-isospin SU(2)$_R$
symmetry, while U(1)$_V$ is a new Abelian invariance intrinsic to
q-$\ell$ symmetric models.
Under this slightly extended gauge group, the fermion field
representations are,
\begin{eqnarray}
&F_L \sim (3,1,2)(0,-1),\quad E_R \sim (3,1,1)(-1,-1),\quad N_R \sim
(3,1,1)(1,-1),&\ \nonumber\\
&Q_L \sim (1,3,2)(0,1),\quad d_R \sim (1,3,1)(-1,1),\quad u_R \sim
(1,3,1)(1,1),&\
\label{g6fermions}
\end{eqnarray}
where $F_L$, $E_R$ and $N_R$ are generalizations of the usual lepton
fields $f_L \equiv (\nu_L,\ e_L)^{\top}$, $e_R$ and $\nu_R$
respectively. The generator $X$ of Eq.~(\ref{Gql}) is given by
\begin{equation}
X = R + V/3,
\end{equation}
while, as for the minimal q-$\ell$ symmetric model,
weak-hypercharge is given by
\begin{equation}
Y = X + T/3,
\label{Y}
\end{equation}
where $T \equiv {\rm diag}(-2,1,1)$ in leptonic colour space. As before,
the formula for $Y$ identifies the standard
leptons as the $T=-2$ components
of the SU(3)$_{\ell}$ triplet fermions, while the $T=1$ components are
the exotic charge $\pm1/2$ liptons.

Many different types of discrete q-$\ell$ symmetries may be defined for
the fermion spectrum of Eq.~(\ref{g6fermions}). We will consider the one
which is defined by the transformations,
\begin{eqnarray}
&F_L \leftrightarrow Q_L,\quad E_R \leftrightarrow d_R,\quad N_R
\leftrightarrow u_R,&\ \nonumber\\
&G^{\mu}_q \leftrightarrow G^{\mu}_{\ell},\quad W^{\mu}
\leftrightarrow W^{\mu},\quad R^{\mu} \leftrightarrow R^{\mu},
\quad V^{\mu} \leftrightarrow -V^{\mu},&\
\label{qlsym}
\end{eqnarray}
where $G^{\mu}_{q,\ell}$ are quark-like and leptonic gluons respectively,
$W^{\mu}$ are weak bosons, and $R^{\mu}$ and $V^{\mu}$ are
the gauge boson fields of U(1)$_R$ and U(1)$_V$ respectively. Note that
this discrete q-$\ell$ symmetry is different from the one in the minimal
model [see Eq.(\ref{minimalqlsym})]. It is important to also realise that
any q-$\ell$ symmetry may be modified by specifying a relative
phase change for the quark and lepton fields when they interchange.
The model specified by Eqs.~(\ref{Gqlpr}-\ref{qlsym}) has not been
explicitly analysed before in the literature. However, a close cousin of
it is discussed in Sec.IIIB of Ref.~{\cite{flvph}}.

The gauge group $G'_{q\ell}$ and a phase-transformed
version of the discrete symmetry
given by Eq.~(\ref{qlsym}) can be simultaneously embedded in a larger
continuous symmetry. The new gauge group is given by $G_6$ where
\begin{equation}
G_6 = SU(6)_{PS} \otimes SU(2)_L \otimes U(1)_R,
\end{equation}
where the subscript ``PS'' refers to Pati-Salam\cite{ps}.
The quarks and their corresponding generalized leptons are
placed in the same multiplet under $G_6$. The fermion multiplet structure
is in fact,
\begin{equation}
\psi_L \sim (6,2)(0),\quad \psi_{1R} \sim (6,1)(1),\quad \psi_{2R} \sim
(6,1)(-1),
\label{G6fermions}
\end{equation}
where $F_L$ and $Q_L$ are inside $\psi_L$, $N_R$ and $u_R$ are inside
$\psi_{1R}$, and $E_R$ and $d_R$ are inside $\psi_{2R}$. If we write the
sextets as column matrices, then we will identify the quark colours
with the upper three components while the lower three components will be
the generalized leptons. The charge $V$ is now the diagonal
generator of SU(6)$_{PS}$ which is given by ${\rm diag}({\bf 1}, -{\bf
1})$ where ${\bf 1}$ is the $3 \times 3$ unit matrix.

How is the discrete quark-lepton symmetry embedded in SU(6)$_{PS}$?
The most general matrix which is both an element of the sextet
representation of SU(6) and a quark-lepton interchange operator
is given by ${\cal C}$ where
\begin{equation}
{\cal C} = \left( \begin{array}{c}
\ 0\quad\ \ \ \ \ iD \\
iD^*\quad\ \ \ \ 0
\end{array} \right).
\end{equation}
In this equation,
$D ={\rm diag}\left(e^{i\theta_1}, e^{i\theta_2}, e^{i\theta_3}\right)$
and the phases $\theta_{1,2,3}$ correspond to the most generally allowed
phase transformations of the various quark and lepton colours when they
interchange. The matrix ${\cal C}$ represents the transformation in
Eq.~(\ref{qlsym}) but with a different phase structure. Since
we are not particularly interested in most of these phase
transformations, it is simplest to take $\theta_{1,2,3} = -\pi/2$. The
simplified discrete symmetry matrix is then given by
\begin{equation}
{\cal C} = \left( \begin{array}{c}
\ 0\quad\ \ {\bf 1} \\
-{\bf 1}\quad\ \ 0
\end{array} \right).
\label{C}
\end{equation}
Note that the minus sign in this simplified matrix is necessary to
ensure that ${\cal C}$ has determinant equal to one. Therefore one
cannot escape from complicating the phase structure of Eq.~(\ref{qlsym})
a little. Actually, the discrete symmetry group left over
after SU(6)$_{PS}$ breaking consists of the elements
$\{{\cal I},\ -{\cal I}={\cal C}^2,\ {\cal C},\
{\cal C}^{-1}={\cal C}^{\dagger}\}$ and is isomorphic to $Z_4$, rather
than the $Z_2$ of Eq.~(\ref{qlsym}). The connection with $Z_2$ is
provided by the homomorphism ${\cal I} \to {\cal I}'$, $-{\cal I} \to
{\cal I}'$, ${\cal C} \to {\cal C}'$, ${\cal C}^{-1} \to {\cal C}'$
from $Z_4$ to $Z_2$ where ${\cal I}'$ is the $Z_2$ identity element and
${{\cal C}'}^2 = {\cal I}'$.
This homomorphism identifies those elements of the $Z_4$ symmetry which
are related to each other only by a phase transformation. Note also that
the element ${\cal C}^2 = -{\cal I}$ of $Z_4$ is also the element
$\exp(i\pi V)$ of U(1)$_V$. Therefore all of the purely phase
transforming actions of the $Z_4$ symmetry can be undone by this
U(1)$_V$ gauge transformation.

The idea of embedding quark and lepton colour inside an extended
Pati-Salam symmetry has already been considered
in Ref.~{\cite{flvps}}. In this
previous paper, the full SU(2)$_R$ right-handed weak-isospin group was
postulated, together with an exact discrete left-right symmetry
(parity) which swapped the two weak-isospin sectors. This made the
model possess partial coupling constant unification, with some
attendant constraints on symmetry breaking scales resulting from a
renormalization group analysis of the theory\cite{jv}.
{}From the point of view of
standard cosmology, however, neither the full SU(2)$_R$ symmetry nor the
discrete left-right symmetry should be imposed. Imposition of the
former would lead to a cosmological monopole problem, because the
initial gauge group would not have a U(1) factor,\footnote{Note
that the generator $R$ of the Abelian group factor in $G_6$ contributes
to the formula for electric charge, as given by $Q = I_{3L} + R/2 + V/6
+ T/6$. If the U(1) in $G_6$ had turned out not to contribute to $Q$,
then topologically stable monopoles would have been produced at some
stage in the symmetry breaking process. We will make some more comments
on monopoles later in this section.} while imposition of
the latter would result in its own domain wall problem\cite{vilenkin}.
Unlike its close relative in Ref.~{\cite{flvps}},
the gauged $G_6$ model has no partial coupling
constant unification, and so the symmetry breaking scales are less
constrained.

At the first stage of symmetry breaking we want to break
SU(6)$_{PS}$ down to its SU(3)$_{\ell}\otimes$SU(3)$_q\otimes$U(1)$_V$
subgroup. We
would also prefer to have the discrete q-$\ell$ symmetry -- as given by
${\cal C}$ in Eq.~(\ref{C}) -- remain unbroken
after this initial breakdown of the $G_6$
group. If the discrete symmetry were to break at the same scale as
SU(6)$_{PS}$ then the domain wall problem would be trivially ``solved'',
because the discrete symmetry would have never existed as a
free-standing invariance at any energy scale.\footnote{As a sidelight,
we note that a Higgs boson transforming as a (20,1)(1) multiplet under
$G_6$ can break $G_6$ to the gauge symmetry $G_{q\ell}$ of the minimal
model [see Eq.~(\ref{Gql})] simultaneously with the discrete symmetry.
Therefore, in this case there is definitely no domain wall problem, but
there is also never a free-standing discrete q-$\ell$ symmetry.
Nevertheless, since
the leptonic colour group can remain exact to TeV-scales
even though the discrete q-$\ell$ symmetry might be broken
at a high scale, this scenario is not completely devoid of
phenomenological interest. Note also that a Higgs field in the (35,1)(0)
representation can break the discrete symmetry at the same time as it
induces the breaking of $G_6$ down to $G'_{q\ell}$ (see below).}
We prefer instead to ensure that the effective theory
below the first symmetry breaking scale is a model with an exact,
un-embedded q-$\ell$ symmetry.

This first stage of symmetry breaking can be accomplished in a number of
ways. The simplest method is to introduce
a real Higgs field $\Phi$ whose $G_6$ transformation law is given by
\begin{equation}
\Phi \sim (189,1)(0).
\end{equation}
Under the SU(3)$_{\ell}\otimes$SU(3)$_q\otimes$U(1)$_V$ subgroup of
SU(6)$_{PS}$, the field
\begin{eqnarray}
\Phi & \to & (1,1)(0) \oplus (1,8)(0) \oplus (8,1)(0) \oplus (8,8)(0)
\nonumber \\
\ & \oplus & (3,\bar{3})(-2) \oplus (\bar{3},3)(2)
\oplus (3,\bar{3})(4) \oplus (\bar{3},3)(-4) \nonumber \\
\ & \oplus & (3,6)(-2) \oplus (6,3)(2)
\oplus (\bar{3},\bar{6})(2) \oplus (\bar{6},\bar{3})(-2).
\end{eqnarray}
A non-zero vacuum expectation value (VEV) for the singlet (1,1)(0)
component of $\Phi$ performs the gauge symmetry breaking we require,
this being
\begin{equation}
G_6 \to SU(3)_{\ell} \otimes SU(3)_q \otimes SU(2)_L \otimes U(1)_R
\otimes U(1)_V.
\end{equation}
This breaking is of course just $G_6 \to G'_{q\ell}$. The discrete
q-$\ell$ symmetry is also left unbroken, as we will now show.

The $189$-plet is actually the lowest dimensional representation
one can use to leave the discrete q-$\ell$ symmetry unbroken. The
lower dimensional representations $35$ and $175$ also contain
SU(3)$_{\ell}\otimes$SU(3)$_q\otimes$U(1)$_V$ singlets.
However, VEVs for these components would
break the discrete symmetry, because they are odd under the discrete
symmetry.\footnote{The homomorphism from $Z_4 \to Z_2$
discussed above defines the representation of $Z_4$ under which
SU(3)$_{\ell}\otimes$SU(3)$_q\otimes$U(1)$_V$ singlet component Higgs
fields transform (with ${\cal C}' = -1$).
Since this representation is isomorphic to $Z_2$,
the terms ``odd'' and ``even'' are applicable.}
To see this, consider the
decomposition of the tensor product
\begin{equation}
6 \otimes \overline{6} = 1 \oplus 35
\label{6}
\end{equation}
under the subgroup SU(3)$_{\ell}\otimes$SU(3)$_q\otimes$U(1)$_V$,
where $6 \to (3,1)(-1) \oplus (1,3)(1)$. Denote the two singlets
in the product by $S_1$ and $S_2$; they are given by
\begin{eqnarray}
S_1 & \subset & (1,3)(1) \otimes (1, \overline{3})(-1)\quad
{\rm and} \nonumber \\
S_2 & \subset & (3,1)(-1) \otimes (\overline{3},1)(1).
\end{eqnarray}
Under the operation of the q-$\ell$ symmetry
$S_1 \leftrightarrow S_2$. From these two singlets we can
construct two independent combinations, $S_1 + S_2$ and
$S_1 - S_2$, which transform as even and odd under the
q-$\ell$ symmetry, respectively. By neccessity, the q-$\ell$--even
singlet corresponds to the SU(6) singlet in the right-hand side
of Eq.(\ref{6}).
Therefore the singlet in the $35$-plet must be q-$\ell$--odd.
[One can check this explicitly by using the representation
given in Eq.(\ref{C}).]
By using this result and the same method one can show that the
singlet in the $189$-plet is q-$\ell$--even from the decomposition of
\begin{equation}
15 \otimes \overline{15} = 1 \oplus 35 \oplus 189.
\end{equation}
Similarly, the $175$-plet can be shown to contain a q-$\ell$--odd
singlet by using the decomposition of $20 \otimes 20 = 1 \oplus 35 \oplus
175 \oplus 189$, while the $405$-plet can be shown to contain a
q-$\ell$--even singlet by considering
$\overline{21} \otimes 21 = 1 \oplus 35 \oplus 405$.

The second stage of symmetry breaking is induced through Higgs
multiplets called $\chi$ and $\Delta$ (we require that
$\langle\chi\rangle$, $\langle\Delta\rangle  \ll \langle\Phi\rangle$ in
order to create the possibility of interesting TeV-scale phenomenology).
These fields possesses Yukawa couplings to the
fermions given by the Lagrangian ${\cal L}_{\rm Yuk}$ where
\begin{equation}
{\cal L}_{\rm Yuk}  =  h_L \bar{\psi}_L \chi (\psi_L)^c + h_R
\bar{\psi}_{1R} \chi (\psi_{2R})^c \nonumber\\
 +  n \bar{\psi}_{1R} \Delta (\psi_{1R})^c + {\rm H.c.}\ ,
\label{Yuk}
\end{equation}
and their transformation properties under $G_6$ are,
\begin{equation}
\chi \sim (15,1)(0)\qquad {\rm and}\qquad \Delta \sim (21,1)(2).
\end{equation}
Since $G_6 \to G'_{q\ell}$ at the first stage of symmetry breaking, we also
need to know how $\chi$ and $\Delta$ transform under the
unbroken subgroup. The branching rules to $G'_{q\ell}$ are
\begin{eqnarray}
\chi & \to & (\bar{3},1,1)(0,-2) \oplus (1,\bar{3},1)(0,2) \oplus
(3,3,1)(0,0);\nonumber\\
\chi & \to & \chi_1 \oplus \chi_2 \oplus \chi',
\end{eqnarray}
and
\begin{eqnarray}
\Delta & \to & (6,1,1)(2,-2) \oplus (1,6,1)(2,2) \oplus (3,3,1)(2,0);
\nonumber\\
\Delta & \to & \Delta_1 \oplus \Delta_2 \oplus \Delta',
\end{eqnarray}
where the equation below each branching rule establishes our notation
for the multiplets which are irreducible under the unbroken gauge group.
Clearly, under the discrete q-$\ell$ symmetry (ignoring the phases)
\begin{equation}
\chi_1 \leftrightarrow \chi_2\qquad {\rm and}\qquad \Delta_1
\leftrightarrow \Delta_2,
\end{equation}
while the components of $\chi'$ and $\Delta'$ transform amongst
themselves. The Higgs fields $\chi_{1,2}$ and $\Delta_{1,2}$ correspond to
their namesakes in the minimal q-$\ell$ symmetric model reviewed in Sec.3.

The multiplets $\chi_1$ and $\Delta_2$
can be represented by antisymmetric and
symmetric $3 \times 3$ matrices, respectively. Under SU(3)$_{\ell}$
transformations
\begin{equation}
\chi_1 \to U_{\ell} \chi_1 U_{\ell}^{\top}\qquad {\rm and}\qquad
\Delta_1 \to U_{\ell} \Delta_1 U_{\ell}^{\top},
\end{equation}
where $U_{\ell}$ is a triplet representation matrix of
an SU(3)$_{\ell}$ group element.
We require the $\chi_1$ and $\Delta_1$
components of the full Higgs multiplets $\chi$ and $\Delta$
to develop nonzero VEVs in order to break both the discrete q-$\ell$
symmetry and the leptonic colour group SU(3)$_{\ell}$. Of course, the
other multiplets inside $\chi$ and $\Delta$ possess quark colour and so we
must demand that their VEVs be zero. The required pattern of VEVs is
\begin{equation}
\langle\chi_1\rangle = \left( \begin{array}{c}
0\ \quad \ 0\quad 0 \\ 0\ \quad \ 0\quad v \\ 0\quad -v\quad 0
\end{array} \right)
\end{equation}
and
\begin{equation}
\langle \Delta_1\rangle = \left( \begin{array}{c}
v'\quad\ 0\quad 0 \\ 0\ \quad\ 0\quad 0 \\ 0\ \quad\ 0\quad 0
\end{array} \right),
\end{equation}
with $\chi_2$, $\chi'$, $\Delta_2$ and $\Delta'$ all having zero VEVs.

It is important to note that the trilinear term
$\chi^{\dagger}\chi\Phi$ appears in the Higgs potential for this model.
This term ensures that the discrete transformation $\Phi \to -\Phi$ is
not an accidental symmetry of the theory, and so there is no accidental
domain wall problem either. One can check that there are no other
accidental discrete symmetries in the model that are not also elements
of a continuous global symmetry. This term
also serves to connect the $\Phi$ multiplet in
a non-trivial way with the other Higgs fields of the theory.

After the second stage of symmetry breaking, the unbroken gauge group is
SU(2)$'\otimes{G_{SM}}$, where $G_{SM}$ is just the Standard Model group
SU(3)$_q\otimes$ SU(2)$_L\otimes$U(1)$_Y$ [with $Y$ given by
Eq.~(\ref{Y})] and SU(2)$'$ is an
unbroken remnant of leptonic colour. The
exotic partners of the leptons (the liptons) gain mass from
the $h_{L,R}$ terms in Eq.~(\ref{Yuk}), while the right-handed neutrinos
develop Majorana masses from the $n$ term in the Yukawa Lagrangian. All
of these masses are thus expected to be heavy compared with the usual
leptons and quarks (which are still massless). The heavy charge $\pm$1/2
lipton fields are doublets under the unbroken shard SU(2)$'$. If the
number of fermion generations is not too large (for instance, if there
are three of them), then SU(2)$'$ is asymptotically-free and thus is
expected to be confining. All particles which have non-trivial quantum
numbers under SU(2)$'$ (liptons, some Higgs bosons and some heavy gauge
bosons) are then confined into unstable, integrally-charged bound states.
This neatly evades a potential cosmological abundance problem, because
the lightest half-integrally charged particle would be stable if it
were free.\footnote{Strictly speaking, this particle is stable even if
it is confined. However, in this case its stability is not a problem
for the same reason that there are no stable mesons.} Finally,
note that a large Majorana mass for the right-handed neutrinos sets the
stage for the see-saw mechanism\cite{seesaw}
once electroweak symmetry is broken.

The final stage of symmetry breaking just involves
the usual spontaneous violation of the electroweak group. This is
performed
in the standard way through electroweak Higgs doublets, which also
induce masses for quarks and the usual leptons (we of course require
that $\langle\phi\rangle \ll \langle\chi_1\rangle,
\langle\Delta_1\rangle$). If only one doublet is
used, then there are quark-lepton mass relations at tree-level
of the form $m_u =
m^{\rm Dirac}_{\nu}$ and $m_d = m_e$ due to the discrete q-$\ell$
symmetry. Note that these mass relations are different from those
obtained in the minimal q-$\ell$ symmetric model [see
Eq.~(\ref{massrels})]. Because of this, radiative
corrections in the model that break the tree-level relations can yield
correct but unpredictive quark and lepton masses\cite{flmass}.
Also, if more than one
doublet is used, these mass relations no
longer hold at tree-level (and predictivity is also unfortunately lost).

This essentially completes our demonstration
that the domain wall problem for
spontaneously broken discrete q-$\ell$ symmetry can be evaded by
embedding the discrete transformation in a continuous gauge group.
However, the attentive reader may have noticed a complication arising
with regard to monopoles because of the way we have performed the
spontaneous symmetry breaking. This issue requires some
further discussion:

The point is that the first stage of symmetry
breaking consists of SU(6)$_{PS}$
$\to$ SU(3)$_{\ell}\otimes$SU(3)$_q\otimes$U(1)$_V$, with no
participation from
SU(2)$_L\otimes$U(1)$_R$. The fact that U(1)$_V$ comes
entirely from SU(6)$_{PS}$ means that monopoles exhibiting $V$-type
magnetic charge will be created during the first phase
transition.\footnote{To be more precise, the global structure of the
unbroken group is actually
SU(3)$_{\ell}\otimes$SU(3)$_q\otimes$U(1)$_V$/$Z_3$ so the monopoles
also carry some non-abelian magnetic charge.}
Indeed, if no further symmetry breaking were to take place, these monopoles
would be topologically stable. However, we know that after the final
stage of symmetry breaking $G_6$ has broken to
SU(2)$'\otimes$SU(3)$_q\otimes$U(1)$_Q$ where the generator $R$
contributes to $Q$ (see footnote 11). Since this
breaking cannot support topologically stable monopoles, the monopole-like
states produced at the first stage of symmetry breaking must disappear
in some manner. Although a detailed analysis of how this occurs is well
beyond the scope of this paper, we can fairly easily identify at least
two important processes. First, since they are not topologically
stable once all of the symmetry breaking is complete,
it must be true that the monopoles can just decay into ordinary
forms of energy. Second, at some point after the first phase transition
we have to break a U(1) gauge group, which should lead to the formation
of cosmic strings [which are different from the cosmic strings produced
when the Pati-Salam SU(6) breaks to the discrete symmetry].
Since the generator $V$ contributes to the
generator of this broken U(1), we would expect these cosmic strings to
end in monopoles and antimonopoles, so enhancing their annihilation
rate\cite{monopole}.
We therefore conclude that although monopole-like states will
exist during a certain epoch in the early universe, they will ultimately
disappear and thus in all probability not cause any cosmological
problems.\footnote{If the U(1)$_V$
symmetry never exists as an exact symmetry in its un-embedded form, then
of course no monopoles, unstable or otherwise, ever form. This will be
true, for instance, if $\langle\Phi\rangle \sim \langle\chi_1\rangle$ or
$\langle\Phi\rangle \sim \langle\Delta_1\rangle$ (of course in this case
the discrete q-$\ell$ symmetry would also not exist as a free-standing
invariance). The requirement that
the monopoles disappear quickly enough to be cosmologically acceptable
therefore translates into an upper bound on $|\langle\Phi\rangle -
\langle\chi_1\rangle|$ or
$|\langle\Phi\rangle - \langle\Delta_1\rangle|$. A detailed dynamical
calculation would be necessary to determine this bound, but we expect
that reasonable values for the VEVs would be allowed.}

In order to round off the discussion, we will now briefly address
some further issues:
(i) There are many phenomenological constraints one could place on this
model. We will not derive any bounds here, because we do not want
to obscure our essential point about how the domain wall problem can be
avoided. (ii) The scale at which SU(2)$'$ confines is approximately
calculable, because the leptonic colour coupling constant is equal to
the strong coupling constant at the scale of q-$\ell$ symmetry breaking.
If the q-$\ell$ symmetry breaking scale is not much higher than the
lipton mass scale,
then the confinement energy turns out to be about the same
as for QCD. If there is a splitting between the discrete symmetry
breaking scale and the lipton mass scale (i.e. if
$\langle\Delta_1\rangle \gg \langle\chi_1\rangle$), then the SU(2)$'$
confinement energy is lower than its QCD counterpart.
(iii) That the SU(2)$'$ confinement scale is of the order of hundreds
of MeV or lower implies that the
lowest mass survivors from the underlying q-$\ell$ symmetric model are
the SU(2)$'$ glueballs. These objects may give rise to interesting
phenomenology, and they are also of potential cosmological significance
because they are long-lived. If these glueballs are very light ($\sim 1$
keV), then it has been shown that they do not interfere with standard
Big Bang nucleosynthesis, and that they are a dark matter
candidate\cite{flvps}. If, on the other hand, the glueballs have masses in
the 1 GeV range, then they have to decay in less than about 1 second
in order to be compatible with standard BBN.
Although a detailed analysis of glueballs in this mass range
has not as yet been
carried out for q-$\ell$ symmetric models, a brief study was made in
Ref.\cite{flvglue} which suggested that a range of parameters for the
model allowing the glueballs to be cosmologically acceptable
exists\cite{su5glueballs}.

\vskip 1cm
\leftline{\bf 5. Spontaneous discrete symmetry breaking and the}
\leftline{\bf \ \ \ \ cosmological phase transition}
\vskip 5mm

In this section we will examine whether or not there is {\it necessarily}
a cosmological phase transition associated with the spontaneous
breaking of the discrete symmetry\cite{rnmgs}.
If no such phase transition
need exist (i.e. if no symmetry restoration need occur at some critical
temperature $T_c$), then one can consistently
attribute the broken symmetry as a special initial condition
of the Big Bang. If this is the case then we can arrange for the
vacua in casually disconnected regions to be the same.
Hence the formation of domain walls is avoided. Of course, this very
special initial condition would ultimately require a deep explanation.
However, for our present purposes it is not necessary to push the
analysis to this extreme, given our overwhelming ignorance of physics at
the Planck scale\cite{turok}.

Before proceeding, note that we are assuming that it is
fundamental Higgs scalars which
are responsible for the origin of spontaneous symmetry breaking.
However, the Higgs sector of the SM is experimentally untested
so the origin of spontaneous symmetry breaking remains unclear.
It could well be that spontaneous symmetry breaking is dynamical in
origin and has nothing to do with fundamental scalars\cite{dsb}.
If this is the case then it is still an open question as to whether
symmetry restoration occurs at high temperatures\cite{chang}.

Consider the zero temperature Higgs potential of the minimal
q-$\ell$ symmetric model given by
\begin{eqnarray}
V_0 & = & \lambda_1
\left[ \chi_1^\dagger\chi_1 + \chi_2^\dagger\chi_2 -v^2 \right]^2
\nonumber\\
\ & + & \lambda_2 \chi_1^\dagger\chi_1 \chi_2^\dagger\chi_2
 + \lambda_3 \left( \phi^\dagger\phi - u^2 \right)^2
\nonumber\\
\ & + & \lambda_4
\left[ \phi^\dagger\phi - u^2 + \chi_1^\dagger\chi_1
+ \chi_2^\dagger\chi_2 -v^2 \right]^2.
\label{v1}
\end{eqnarray}
(For illustrative purposes, we have kept the Higgs potential simple by
not including the fields $\Delta_{1,2}$ or multiple copies of $\phi$.)
If the coefficients, $\lambda_i$, are all positive then
the above potential is minimised where $u$ and $v$ are
the nonzero VEVs of the $\phi$ and $\chi_1$ fields respectively.
This then leads to the symmetry breaking pattern given in Eq.~(\ref{pat}).
As a result of this symmetry breaking, there will be two residual
neutral Higgs bosons (coming from $\phi$ and $\chi_1$) whose
mass (squared) matrix is given by
\begin{equation}
\left( \begin{array}{c}
4(\lambda_3+\lambda_4)u^2 \ \quad\quad\quad\ 4\lambda_4uv \\
4\lambda_4uv\ \quad\quad\quad \ 4(\lambda_1+\lambda_4)v^2
\end{array} \right).
\label{HM}
\end{equation}
There will also be the charge 1/3 colour triplet Higgs multiplet
$\chi_2$ with mass given by $M_{\chi_2}^2 = \lambda_2 v^2$.

For the purposes of this section we will rewrite Eq.~(\ref{v1})
in a more convenient form as follows:
\begin{eqnarray}
V_0 & = & -\mu_\phi^2 \phi^\dagger\phi
+(\lambda_3 + \lambda_4 )\left(\phi^\dagger\phi\right)^2
\nonumber\\
\ & + & 2\lambda_4 \left(\phi^\dagger\phi\right)
\left[\chi_1^\dagger\chi_1 + \chi_2^\dagger\chi_2 \right]
\nonumber\\
\ & + & \left(2\lambda_1 + 2\lambda_4 + \lambda_2 \right)
\chi_1^\dagger\chi_1 \chi_2^\dagger\chi_2
\nonumber\\
\ & - & \mu_\chi^2 \left[\chi_1^\dagger\chi_1 + \chi_2^\dagger\chi_2 \right]
\nonumber\\
\ & + & (\lambda_1 + \lambda_4)
\left[\left(\chi_1^\dagger\chi_1\right)^2
+ \left(\chi_2^\dagger\chi_2\right)^2 \right].
\label{v2}
\end{eqnarray}
The minimisation conditions then become
\begin{eqnarray}
ku^2 & = & (\lambda_1 + \lambda_4 )\mu_\phi^2 - \lambda_4\mu_\chi^2,
\nonumber\\
kv^2 & = & (\lambda_3 + \lambda_4 )\mu_\chi^2 - \lambda_4\mu_\phi^2,
\label{min1}
\end{eqnarray}
where $k \equiv 2\left(\lambda_1\lambda_3 + \lambda_1\lambda_4
+ \lambda_3\lambda_4 \right)$.

Now consider the finite temperature contributions to the
effective Higgs potential.
We will only consider the terms proportional to $T^2$
since it is sufficient for us to work within the high temperature
expansion approximation.
By using the usual calculational techniques\cite{dolan},
the finite temperature corrections\footnote{In our high T
approximation we have neglected the contributions that are
proportional to T and the logarithmic corrections to the
coefficients of the quartic terms in the potential.}
modify the minimisation conditions of Eq.~(\ref{min1})
to become
\begin{eqnarray}
ku^2 & = & \left(\lambda_1 + \lambda_4 \right)
\left(\mu_\phi^2 - \zeta_1 T^2 \right)
- \lambda_4\left(\mu_\chi^2 - \zeta_2 T^2 \right),
\nonumber\\
kv^2 & = & \left(\lambda_3 + \lambda_4 \right)
\left(\mu_\chi^2 - \zeta_2 T^2 \right)
- \lambda_4 \left(\mu_\phi^2 - \zeta_1 T^2 \right),
\label{min2}
\end{eqnarray}
where
\begin{eqnarray}
\zeta_1 & = & {1\over 2} \lambda_3 + {3\over 2} \lambda_4
+ {3\over 16} g_2^2 + {1\over 16} g_X^2 + {1\over 2} \Gamma_t^2,
\nonumber\\
\zeta_2 & = & {7\over 6} \lambda_1 + {1\over 4} \lambda_2
+ {3\over 2} \lambda_4 + {1\over 3} g_3^2 + {1\over 36} g_X^2,
\end{eqnarray}
and $g_{X, 2, 3}$ are the $U(1)_X$, $SU(2)_L$, $SU(3)_{q, \ell}$
gauge coupling constants respectively and $\Gamma_t$ is the t-quark
Yukawa coupling constant. (Note that our theory has two Yukawa coupling
constants equal to $\Gamma_t$ because of the discrete symmetry.)
To simplify Eq.~(\ref{min2}) let
$\lambda_3 \ll \lambda_4 $ and $\mu \equiv \mu_\phi \simeq \mu_\chi$.
Then
\begin{eqnarray}
ku^2 & \simeq & \lambda_1 \mu^2 - \lambda_4 \zeta T^2,
\nonumber\\
kv^2 & \simeq & \lambda_3 \mu^2 + \lambda_4 \zeta T^2,
\label{min3}
\end{eqnarray}
where
\begin{equation}
\zeta  =  {1\over 2} \lambda_3 - {7\over 6} \lambda_1 - {1\over 4} \lambda_2
+ {1\over 2} \Gamma_t^2 - {1\over 3} g_3^2
+ {3\over 16} g_2^2 + {5\over 144} g_X^2.
\end{equation}
Therefore $v^2$ can remain nonzero, and hence q-$\ell$ symmetry
unbroken, provided $\zeta \ge 0$.  Clearly, a range of parameters
exists for which this is true.\footnote{Note,
however, that the electroweak
phase transition is still expected to take place. This is because for
temperatures lower than the mass of the lightest exotic particle (be it
a lipton or an exotic Higgs boson or whatever), the effect of all these
non-standard states is Boltzmann suppressed, and so the effective
finite-temperature field theory is essentially that of the SM. The
nature of the electroweak phase transition may be altered because the
lightest Higgs mass eigenstate may have different properties from the
standard Higgs boson and because of the possiblity that one or more of
the exotic particles may fortuitously have masses as low as, say, 100
GeV. However, interesting though they may be, these details are
unimportant for our present purpose.}
More specifically we can choose, for example,
\begin{eqnarray}
\lambda_3 & \ge & {7\over 3}\lambda_1 + {1\over 2}\lambda_2,
\nonumber \\
\Gamma_t^2 & \ge & {2\over 3}g_3^2 - {3\over 8}g_2^2 - {5\over 72}g_X^2,
\end{eqnarray}
where the couplings are evaluated at high T. For such a range of
parameters, the neutral Higgs boson masses at zero temperature
from Eq.(\ref{HM}) are given by
\begin{eqnarray}
M_\phi^2 & \simeq & 4(\lambda_1 + \lambda_3 )
{u^2 \over 1+ {u^2\over v^2}}, \nonumber \\
M_{\chi_1}^2 & \simeq & 4\lambda_4 \left( u^2 + v^2 \right).
\end{eqnarray}
For a Higgs boson mass greater than about 50 GeV (the current
lower limit is 48 GeV \cite{Decamp}) gives
$\lambda_1 + \lambda_3 \ge 0.02$ which is consistent with the
chosen range of parameters in our example.

The Higgs sector we analyzed above is of course unrealistic from the
point of view of fermion mass relations (see Sec.3). We have, however,
checked that a realistic theory containing two electroweak Higgs
doublets yields the same qualitative conclusions as we reached in our
simple illustrative model. For reasons of clarity we have therefore
chosen to explicitly display only the simplified analysis.

So the minimal q-$\ell$ symmetric model has the necessary ingredients
to prevent a restoration of q-$\ell$ discrete symmetry at high temperatures
(and, for that matter, a restoration of leptonic colour symmetry).
Clearly, in extensions of the minimal model, where the
Higgs sector will generally be more complicated, this will also be the case
(the two electroweak doublet extension alluded to in the previous
paragraph is an example).
Such a scenario provides one way of evading the domain wall problem.
It is interesting to also note that electroweak symmetry can be
restored even though the q-$\ell$ symmetry remains broken. This may
prove useful for baryogenesis at the electroweak scale.

\vskip 1cm
\newpage
\leftline{\bf 6. Domain walls and inflation}
\vskip 5mm

As we discussed in Sec.2, the unadorned Hot Big Bang Model cannot
account for the smoothness and flatness of our universe (although it is
compatible with it). The interesting idea of ``inflation'' has been much
studied as a way of remedying this deficiency\cite{linde}.
At some very early stage
in the evolution of the universe, a finite period of exponential
expansion is postulated, which renders spacetime
almost perfectly flat after the exponential expansion ceases.
Also, the present observable universe
arises from within a causally-connected region of the very early
universe, thus explaining its palpable smoothness.

Since its inception, inflation has also served to rid the universe of
otherwise troubling topological defects, provided the period of
inflation occurs after the cosmological phase transition that creates
the topological defects. For instance, one of the original motivations
for inflation was to cure grand unified theories (GUTs) of their monopole
abundance problem. The cure is so efficacious, in fact, that from the
pre-inflation prediction that GUT monopoles dominate the energy density
of the universe by many orders of magnitude, the universe observable to
us today after inflation is predicted to contain at most one
monopole.

As for monopoles, an inflationary epoch after a cosmological phase
transition associated with spontaneous discrete symmetry breaking
also eliminates domain walls from the observable universe. Clearly,
therefore, domain walls generated by discrete q-$\ell$ symmetry can be
rendered innocuous by this means.

The only issue we have to really discuss in this regard is the
relative positioning of the scales
of symmetry breaking in q-$\ell$ symmetric models and the
scale $\Lambda_{\rm Inf}$ at which inflation occurs.\footnote{For
inflation to solve the smoothness and flatness problems, the period of
exponential expansion must be sufficiently long. The scale $\Lambda_{\rm
Inf}$ is then to be interpreted as corresponding to the temperature at
which inflation begins. During the inflationary phase, the universe
supercools so that $\Lambda_{\rm Inf}$ no longer even approximately
corresponds to the temperature of the universe at that time. When
inflation ceases the universe is reheated by the conversion of false
vacuum energy into thermal energy for the particle soup. The reheating
temperature turns out to be less than $\Lambda_{\rm Inf}$
so inflation does not restart and the discrete
q-$\ell$ symmetry is not restored.}
In general, we
would expect the inflationary phase transition to occur at a very high
scale, say about $10^{14}$ GeV. This means, for a start,
that we must arrange
spontaneous q-$\ell$ symmetry breaking to occur at an even higher
scale.

Since we would like a lot of new phenomenology to occur in the
TeV energy regime, we may like to consider divorcing
the discrete symmetry breaking
scale $\Lambda_{q\ell}$ from the leptonic colour breaking scale
$\Lambda_3$ and/or the lipton mass scale $\Lambda_L$.
It is sensible, in fact, to consider three hierarchical patterns:
\begin{equation}
\Lambda_{q\ell} > \Lambda_{\rm Inf} \gg \Lambda_3 \sim \Lambda_L
\label{hier1}
\end{equation}
or
\begin{equation}
\Lambda_{q\ell} > \Lambda_{\rm Inf} \gg \Lambda_3 \gg \Lambda_L
\label{hier2}
\end{equation}
or
\begin{equation}
\Lambda_{q\ell} \sim \Lambda_3 > \Lambda_{\rm Inf} \gg \Lambda_L.
\label{hier3}
\end{equation}
Let us begin thinking about these patterns in terms of the minimal
q-$\ell$ symmetric model introduced in the Sec.3.\footnote{Generically,
we would expect the scales $\Lambda_{3,L}$ to roughly correspond
to the temperatures at which the associated cosmological phase
transitions occur. Note, however, that this need not be true for reasons
outlined in the previous section. Note also that if the reheating
temperature after inflation is lower than either
$\Lambda_3$ or $\Lambda_L$ or both, then the associated phase
transition(s) will not occur in the post-inflationary universe.}
(Do not worry, for the near future, about how the scale
of inflation $\Lambda_{\rm Inf}$ is to be
generated. We will just assume in an ad-hoc way that an inflaton field
can be added to the model to bring about inflation at any desired scale.
Fitting the inflaton field into the rest of particle physics in an
elegant way is a very deep and unsolved problem which we are not going
to address.) It is immediately apparent that we have to extend the Higgs
sector of the minimal q-$\ell$ symmetric model in order to generate the
hierarchies of Eqs.~(\ref{hier1}) and (\ref{hier2}) (call them Hierarchy
1 and Hierarchy 2 respectively). This is because the Higgs fields
$\chi_1$ and $\Delta_1$ introduced in the Sec.3 both
simultaneously break the discrete symmetry and leptonic colour. Thus we
need to introduce another Higgs field $\sigma$ which is a gauge
singlet but which is odd under q-$\ell$ symmetry. Note, however, that
$\chi_1$ and $\Delta_1$ are sufficient in order to generate Hierarchy 3
[see Eq.~(\ref{hier3})].

In terms of VEVs for Higgs fields,
\begin{equation}
\Lambda_{q\ell} \sim {\rm max}(\langle\sigma\rangle,
\langle\chi_1\rangle,\langle\Delta_1\rangle),\quad
\Lambda_L \sim \langle\chi_1\rangle\quad {\rm and}\quad
\Lambda_3 \sim {\rm max}(\langle\chi_1\rangle,\langle\Delta_1\rangle).
\end{equation}
Hierarchies 1, 2 and 3 are generated, respectively, if
\begin{equation}
\langle\sigma\rangle > \Lambda_{\rm Inf} \gg \langle\Delta_1\rangle
\sim \langle\chi_1\rangle,
\end{equation}
or
\begin{equation}
\langle\sigma\rangle > \Lambda_{\rm Inf} \gg \langle\Delta_1\rangle
\gg \langle\chi_1\rangle,
\end{equation}
or
\begin{equation}
\langle\Delta_1\rangle > \Lambda_{\rm Inf} \gg \langle\chi_1\rangle.
\end{equation}
We should be aware that some fine-tuning of parameters will be
necessary in order to generate these hierarchies. Since the purpose of
the present paper is to show how domain walls can be made cosmologically
safe, we do not want to cloud the issue by including complicated
speculations about how the gauge hierarchy problem might eventually be
alleviated. It is sufficient for us that such hierarchies
{\it can} be induced.

We do not need to discuss Hierarchy 3 much further. We simply fine-tune
the Higgs potential parameters to create this hierarchy, and we throw in
an inflaton field. Note also that a fine-tuning is necessary to
keep the lipton masses light after leptonic colour is broken, because
the gauge group SU(2)$' \otimes G_{SM}$ cannot by itself
prevent the radiative generation of nonzero lipton masses. If we extend
the minimal q-$\ell$ symmetric model gauge group, then it is possible to
have a symmetry left over after the first stage of symmetry breaking in
Hierarchy 3 which does prevent the liptons from
gaining mass\cite{flvph}.

Hierarchies 1 and 2 require the additional Higgs field $\sigma$. This is
a real Higgs field which is odd under discrete q-$\ell$ symmetry. It
couples to the other fields in the model through the Higgs potential
terms,
\begin{eqnarray}
V_{\sigma} & = & -\mu^2_{\sigma} \sigma^2 + \lambda_{\sigma} \sigma^4 +
a_{\chi}(\chi^{\dagger}_1 \chi_1 - \chi^{\dagger}_2 \chi_2) \sigma +
a_{\Delta}(\Delta^{\dagger}_1 \Delta_1 - \Delta^{\dagger}_2 \Delta_2)
\sigma \nonumber\\
& + & \lambda_{\sigma\chi} (\chi^{\dagger}_1 \chi_1 + \chi^{\dagger}_2
\chi_2) \sigma^2 + \lambda_{\sigma\Delta} (\Delta^{\dagger}_1 \Delta_1 +
\Delta^{\dagger}_2 \Delta_2) \sigma^2 + V(\phi,\sigma),
\end{eqnarray}
where $V(\phi,\sigma)$ describes the coupling of $\sigma$ to however
many electroweak doublets $\phi$ we have in our theory.
The two trilinear terms in this equation
establish $\sigma$'s q-$\ell$--odd credentials. Again,
we fine-tune the parameters in the full Higgs potential
in order to generate either Hierarchy 1 or Hierarchy 2. Note that after
the discrete symmetry is spontaneously broken, the coupling constants of
the two colour forces will evolve a little
differently under the renormalization
group, due to the fact that the leptonically-coloured Higgs fields will
now have different masses from those with quark colour.

So, we conclude this section by saying that it is possible to
inflate-away the domain walls from q-$\ell$ symmetry breaking, while
also preserving the feature of having new low-energy phenomenology. This
new phenomenology can be either the existence of light liptons (say 100
GeV and above), or the existence of both light liptons and all the new
physics associated with the breaking of leptonic colour (Higgs fields
and heavy gauge bosons).

\vskip 1cm
\leftline{\bf 7. Lifting the vacuum degeneracy}
\vskip 5mm

When a $Z_2$ discrete symmetry spontaneously breaks,
the standard perturbative
analysis of the Higgs potential reveals a vacuum manifold
consisting of two degenerate states that can be transformed into each
other. An exact degeneracy is necessary for the resulting domain walls
to be completely stable.

If a perturbation is added to the theory that explicitly breaks
the discrete symmetry, then these two states are no longer exactly
degenerate. Provided the explicit breaking is small enough, domain wall
structures can still form, but they will no longer be stable. Since it
is energetically favoured for the true vacuum state to be established
throughout the universe, these domain walls have to break up eventually.
One can view this process as being caused by a pressure differential
across the domain wall, due to the slightly different energy densities
on each side.

A ``cheap and nasty'' way out of the domain wall problem for q-$\ell$
symmetric models is therefore to include a small amount of explicit
breaking. One can even be so sophisticated as to include only soft
breaking terms. One would also have to be careful to make the explicit
breaking strong enough so that the domain walls break up quickly enough.
However, we view such models as unpalatable since they
render the term ``quark-lepton symmetry'' a misnomer.

A somewhat more attractive possibility exists, however, for it
could turn out that {\it non-perturbative effects}
lift the degeneracy. A class of
discrete symmetry models for which this is supposed to occur has
recently been discussed in the literature\cite{preskill}.
They are known as theories with
``anomalous discrete symmetries''\cite{preskill,ibanez}.

The examples of this phenomenon studied in the literature to date refer
to discrete $Z_n$ symmetries that can be embedded inside the continuous
group U(1) (in a number of different ways, in general) \cite{ibanez,white}.
Such a $Z_n$ discrete symmetry is termed anomalous
if all of the associated U(1) parents are anomalous
with respect to the gauge symmetries of the model. In an interesting
analysis, Preskill et al.\cite{preskill}
have argued that the discrete symmetry imposed
to prevent Higgs-induced tree-level flavour-changing process in the
two-Higgs-doublet model is anomalous, and thus the putative vacuum
degeneracy is lifted by instanton effects.\footnote{An as yet unresolved
controversy exists as to whether anomalous symmetries are ``explicitly''
or ``spontaneously'' broken\cite{christos}.
In the past, this contentious issue has of
course revolved around anomalous continuous symmetries [and in
particular the axial U(1) transformations that are an approximate
symmetry of QCD]. It seems reasonable that a similar uncertainty should
also exist about the status of anomalous discrete symmetries. If it
turns out that anomalous symmetries are to be properly regarded as
explicitly broken, then such transformations are not really symmetries
in the first place. Anomalous q-$\ell$ symmetries -- should they exist
-- would therefore be as misnamed as their
``cheap and nasty'' cousins in the case mentioned above. However, if the
alternative view prevails that anomalous symmetries
are in truth spontaneously broken, then this method of avoiding the
domain wall problem would be rather more attractive. Note that
this point of view requires one to view the
physical consequences of spontaneous breaking differently for
anomalous symmetries compared with those broken in the more conventional
manner. For the axial U(1) of QCD, for instance, Goldstone's Theorem no
longer holds, while for anomalous discrete symmetries the vacuum
degeneracy does not occur. We have nothing new to contribute to this old
debate, but merely wish to alert the uninitiated reader to its
existence.}
They go on to argue that
domain walls caused by this discrete symmetry decay in time to prevent
cosmological difficulties.

Can such a phenomenon also occur for discrete quark-lepton symmetry?
Before addressing this question, we have to generalize the notion of an
anomalous discrete $Z_n$ symmetry to include embeddings inside SU(N)
rather than just U(1). This is because discrete symmetry subgroups of
U(1) act by changing the relative phases of fields, while q-$\ell$
symmetry is an example of a discrete symmetry that {\it interchanges
flavours}. We will call such a group a ``flavour interchanging discrete
symmetry'' or FID for short.

The discussion pertaining to Eq.~(\ref{C}) illustrates how
flavour interchanging embeddings of $Z_4$ inside SU(2N) are
established. To generalize the discussion slightly, consider the
SU(N)$_1 \otimes$ SU(N)$_2 \otimes$ U(1) subgroup of SU(2N). The element
of the fundamental representation of SU(2N) that interchanges the two
SU(N) sectors is given by
\begin{equation} {\cal C}_N = \left( \begin{array}{c}
\ \ \ 0 \ \ \ \ {\bf 1} \\ -{\bf 1}\ \ \ \ 0
\end{array} \right),
\label{CN}
\end{equation}
where ${\bf 1}$ is the $N \times N$ unit matrix (and we have fixed
the phases).
We will call the resulting FID anomaly-free (anomalous) if the
parent SU(2N) gauge theory is anomaly-free (anomalous). This is, of
course, a fairly obvious generalization of the U(1) example studied in
the literature.

In Sec.4 we showed that the discrete symmetry of Eq.~(\ref{qlsym}) could
be embedded within an anomaly-free representation of $G_6 =$
SU(6)$_{PS}\otimes$SU(2)$_L$ $\otimes$U(1)$_R$.
Therefore, this particular version of
q-$\ell$ symmetry is anomaly-free according to our definition. Thus
effects related to anomalies cannot lift the vacuum degeneracy, if the
analysis of Ref.~\cite{preskill} can be validly extended to FIDs (and we
see no obvious reason why it cannot be).

We now consider how the alternative version of q-$\ell$
symmetry given by Eq.~(\ref{minimalqlsym}),
as used in the minimal model, may be
embedded into the $G_6$ model. Since the minimal-model
form of q-$\ell$ symmetry has the interchanges $E_R \leftrightarrow u_R$
and $N_R \leftrightarrow d_R$, it clearly is not an element of
$G_6$. However, if the discrete symmetry ${\cal R}$ given by
\begin{equation}
\psi_{1R} \leftrightarrow \psi_{2R},\quad R^{\mu} \leftrightarrow
-R^{\mu}
\end{equation}
is also imposed, then the discrete symmetry of the minimal model is
given by the diagonal subgroup of ${\cal R} \otimes {\cal C}$ [where ${\cal
C}$ is defined in Eq.~(\ref{C})].

Is this new discrete symmetry ${\cal R}$ anomaly-free or not? It is
trivial to embed (a phase-transformed version of)
this discrete symmetry into an anomaly-free gauge
theory. The symmetry ${\cal R}$ is just a remnant of the
right-handed weak-isospin group SU(2)$_R$. Under the gauge group
SU(6)$_{PS} \otimes$ SU(2)$_L \otimes$ SU(2)$_R$, the fermion
transformation laws are
\begin{equation}
\psi_L \sim (6,2,1),\qquad \psi_R \sim (6,1,2).
\end{equation}
Since this fermion spectrum is anomaly-free, so is ${\cal R}$. Therefore
it follows that the version of quark-lepton symmetry employed in the
minimal model (up to phases) is anomaly-free. [Of course, if the
symmetry SU(6)$_{PS} \otimes$ SU(2)$_L \otimes$ SU(2)$_R$ were actually
gauged, then the model would have a monopole problem, provided the
monopoles were not inflated away, and provided that the full gauge
symmetry were restored at high-temperature.]

There are other versions of q-$\ell$ symmetry that we should also
consider. Take for instance the q-$\ell$ symmetric model that also
features left-right symmetry \cite{flvph,flref1}.
The gauge group is $G_{q\ell LR}$ where
\begin{equation}
G_{q\ell LR} = SU(3)_{\ell} \otimes SU(3)_q \otimes SU(2)_L \otimes
SU(2)_R \otimes U(1)_V,
\end{equation}
under which the fermion classifications are,
\begin{eqnarray}
& F_L \sim (3,1,2,1)(-1),\qquad F_R \sim (3,1,1,2)(-1), &\ \nonumber\\
& Q_L \sim (1,3,2,1)(1),\qquad Q_R \sim (1,3,1,2)(1). &\
\end{eqnarray}
The two q-$\ell$ symmetries we have considered hitherto have involved
interchanging left-handed leptons with left-handed quarks, and
right-handed leptons with right-handed quarks (in two different ways).
But the discrete q-$\ell$ symmetry,
\begin{equation}
F_L \leftrightarrow (Q_R)^c,\qquad F_R \leftrightarrow (Q_L)^c
\label{qlLRsym}
\end{equation}
is also worth looking at. Note that we have not written down
the obvious interchanges of gauge fields necessary to define this
symmetry. In order for this discrete symmetry to be classified either as
anomalous or anomaly-free, we have to find a way of placing $[F_L,
(Q_R)^c]$ and $[F_R, (Q_L)^c]$ into non-abelian gauge group
representations. Such a group would have to be large enough to contain
the whole of $G_{q\ell LR}$ as a subgroup (it would
be a simple GUT group, in
fact). It is clear that there is no such group with the necessary
representations. (Note that we want to do this embedding without
introducing any other fermions into the same GUT multiplets with the
pre-existing leptons and quarks.) We therefore conclude, that the
symmetry of Eq.~(\ref{qlLRsym}) is {\it neither} an anomalous {\it nor}
an anomaly-free discrete symmetry.

So, we cannot use the argument pertaining to anomalous discrete
symmetries to conclude that domain walls are unstable in this model,
even though the discrete symmetry is not anomaly-free. The authors do
not know if there are non-perturbative
effects different from those associated with anomalies
that might lift the vacuum degeneracy in a case such as this.

There is yet one more class of discrete q-$\ell$ symmetry we should
discuss: those also involving the discrete spacetime symmetries of
parity and time-reversal\cite{flvcpt,flql}.
An example will suffice. Consider the minimal
q-$\ell$ symmetric model gauge group $G_{q\ell}$ and its fermion
spectrum Eq.~(\ref{minimalf}). The q-$\ell$ symmetry,
\begin{equation}
F_L \leftrightarrow (Q_L)^c,\quad E_R \leftrightarrow (u_R)^c,\quad
N_R \leftrightarrow (d_R)^c,
\end{equation}
is also a parity symmetry, and requires the spacetime parity
transformation for its consistent definition. (Gauge boson interchanges
are not displayed, and the Lorentz structure is suppressed.) There are
other examples of q-$\ell$ symmetries that are also spacetime
symmetries\cite{flvcpt}.
Clearly these transformations cannot be embedded into any
gauge group representation. Therefore, they are also neither anomalous
nor anomaly-free discrete symmetries. Can any non-perturbative effects
lift the vacuum degeneracies naively implied by spacetime discrete
symmetries? Again, the authors do not know the answer to this question.
(Note that the answer to this question would also be relevant to
the usual discrete parity symmetry of left-right symmetric models,
and to CP transformations, and so on.)

So, we conclude that all q-$\ell$ symmetries are either manifestly
anomaly-free or not embeddable into a gauge group. Models using the
former varieties are expected to have an exact vacuum degeneracy, while
the latter varieties could perhaps do with some further analysis.

\vskip 1cm
\leftline{\bf 8. Conclusion}
\vskip 5mm

We have demonstrated that spontaneously broken discrete quark-lepton
symmetry can be consistent with the standard Hot Big Bang Model of
cosmology. The domain wall problem can be avoided by rendering inoperative
one or more of the usual assumptions made in the standard argument that
domain walls are a cosmological disaster. We found (i) that
domain walls can be made unstable by embedding the discrete
symmetry into a continuous symmetry; (ii) that the
necessary cosmological phase
transition need not occur; and (iii) that
stable domain walls can be inflated-away
even if they form. In each of these scenarios, much interesting new
phenomenology can occur at the TeV scale. The cosmological domain wall
problem therefore does not in any way rule out the possibility of
finding evidence at the TeV scale
for an underlying quark-lepton symmetry in nature.
We also discussed the idea of anomalous discrete
symmetries, but did not find any clear-cut version of quark-lepton
symmetry that could have its vacuum degeneracy lifted by non-perturbative
effects.

\vskip 1cm
\centerline{\bf Acknowledgements}

\noindent
This work was supported in part by a grant from the DOE.
R.R.V. is supported by the Australian Research Council. He would like
to acknowledge discussions with X.-G. He and G. C. Joshi.

\end{document}